\documentclass[a4paper,11pt]{article}
\usepackage{pos}
\usepackage{ulem}

\title{Comparative multi-probe study of jet energy-loss in QGP}

\author*[a]{Rouzbeh Modarresi Yazdi}
\author[b,c]{Shuzhe Shi}
\author[a]{Charles Gale}
\author[a]{Sangyong Jeon}

\affiliation[a]{Department of Physics, McGill University,\\
  3600 Rue University, Montr\'eal, Quebec, Canada, H3A 2T8}
\affiliation[b]{Center for Nuclear Theory, Department of Physics and Astronomy, Stony Brook University\\
Stony Brook, New York 11794–3800, USA}
\affiliation[c]{Department of Physics, Tsinghua University\\
Haidian District, Beijing, 100084, China}

\emailAdd{rouzbeh.modarresi-yazdi@mail.mcgill.ca}
\abstract{
    Jet-energy loss is an important sign of the creation of Quark-Gluon Plasma in heavy-ion collisions. High transverse momentum ($p_T$) partons are produced at the moment of initial hard scattering and are modified as a result of their propagation through the created medium. We study two models of low-virtuality radiative energy loss: \cujet\, and \martini. This is done using the \jetscape\, framework, which allows for an objective comparison. \cujet\, is integrated into the \jetscape\, workflow, and full jet simulations, including substructure observables, are computed for the first time using leading-order \dglv\, rates. Strongly-interacting probes (charged hadrons, jets, jet fragmentation functions and jet shape ratio) are considered along jet-medium photons for the first time in a dynamic QGP. We find that these photons make a significant contribution in the phenomenologically interesting intermediate $p_T$ domain of $4$-$12$ GeV.
}

\FullConference{%
  11th International Conference on Hard and Electromagnetic Probes of High-Energy Nuclear Collisions\\
  26th-31st March 2023\\
  Aschaffenburg, Bavaria, Germany
}

\newcommand{\jetscape}{\textsc{jetscape}}
\newcommand{\martini}{\textsc{martini}}

\newcommand{\trento}{\textsc{t$_{R}$ento}}
\newcommand{\urqmd}{\textsc{UrQMD}}
\newcommand{\cujet}{\textsc{cujet}}
\newcommand{\matter}{\textsc{matter}}
\newcommand{\pythia}{\textsc{pythia}}

\newcommand{\amy}{\textsc{amy}}
\newcommand{\dglv}{\textsc{dglv}}

\newcommand{\pbpb}{\texttt{Pb-Pb}}

\newcommand{\pp}{\texttt{p-p}}
\newcommand{\nuclearcoll}{\texttt{A-A}}

\newcommand{\vishnu}{\textsc{vishnu}}
\newcommand{\raa}{\ensuremath{R_\mathrm{AA}}}

\begin{document}
    \maketitle
    
    \emph{\textbf{Introduction}} ---
    Observation of jet energy loss has been an important signal of the creation of the strongly interacting plasma of quarks and gluons (QGP). In this work we are interested in a comparative study of two important models of parton energy loss in an evolving QGP medium: \cujet\, and \martini. This is done using a composite, multi-stage jet energy-loss model where \matter~\cite{MATTER} models the high-virtuality part of the energy loss and \martini~\cite{martini} or \cujet~\cite{cujet} are used to simulate the low-virtuality part of the evolution. The composite models are constructed using \jetscape~\cite{JETSCAPE:pp,JETSCAPE:other}, a modular framework for jet simulation and energy loss studies. The modularity of \jetscape\, allows for an objective comparative analysis of the models, where all aspects of the evolution except for the one under study, are held fixed.

\vspace{2mm}\emph{\textbf{Parton energy loss}} ---
    In this section we present a brief summary of the physics of the energy loss channels of \martini\, and \cujet. 
    
     \vspace{2mm}
     \emph{Elastic energy loss} --- 
        Both models model elastic scatterings of the hard partons with the medium using $t$-channel gluon exchange. \martini\, uses the HTL-resummed gluon propagator in the evaluation of the matrix elements, while \cujet\, uses the gluon Debye mass as the infrared regulator. If the momentum transfer to the medium parton is large and its momentum is increased above $p_{\mathrm{cut}}=2$ GeV, it is taken into the event record as a hard parton, leaving a hole in the medium. At the end of the evolution, the hole partons are hadronized and their contribution is subtracted from the charged hadrons and jet spectra. 
        
     \vspace{2mm}
     \emph{Radiative energy loss in \cujet} --- 
         In \cujet, the incoming parton is assumed to be propagating through the medium after being created at some finite time within it. The QGP medium is taken to be at very high temperatures, allowing for a weak coupling approximation. The most important assumption in this model is that the QGP medium is thin. Therefore one can expand the radiated gluon spectrum in powers of opacity. Gluon bremsstrahlung in \cujet\, is given by the LO results of \dglv~\cite{dglv} model. The view of the medium constituents is that of a series of well-separated dynamical scattering centers. The density of the scattering centers in the QGP medium, whose value can be determined using the equation of state. A feature of the radiative rates in \cujet\, is their time dependence, a manifestation of the LPM effect, arising from the interference between the different diagrams~\cite{dglv}. 

         A new addition to \cujet\, is photon bremsstrahlung. The rate has been previously computed within the \dglv\, framework to LO in opacity in Ref.~\cite{Zhang:2010hiv} using static scattering centers and a Gaussian profile for their density. We modify that result to include dynamic scattering centers with the same density as gluon bremsstrahlung. The photon rate is then given by 
        \begin{align}
            \begin{split}
           & \frac{\mathrm{d}\Gamma^{\dglv}_{q \to q\gamma}}{\mathrm{d}z}(p,z,\tau)
            =\;
             	\frac{e_f^2\alpha_{\mathrm{em}}}{\pi^2}  \frac{32+8N_f}{16+9N_f} \rho(T)
            	\int{\mathrm{d}^2\mathbf{k}_{\perp}}
            	\frac{(1-z_+)^2}{z_+}\left| \frac{\mathrm{d}z_+}{\mathrm{d}z} \right| 
            	\int \frac{ \mathrm{d}^2\mathbf{q}_{\perp}}{\mathbf{q}_\perp^2} 
            	\frac{ \alpha_s^2(\mathbf{q}_{\perp}^2)}{\mathbf{q}_\perp^2 + m^2_D} \times	
            \\ & 
            	\Bigg[
            	\bigg(\frac{\mathbf{k}'_{\perp}}{{\mathbf{k}_{\perp}'}^2+\chi^2}-
            	\frac{\mathbf{k}_{\perp}}{\mathbf{k}_{\perp}^2+\chi^2} \bigg)^2
            	+ 
            	2\bigg(\frac{\mathbf{k}_{\perp} \cdot \mathbf{k}'_{\perp}}{({\mathbf{k}_{\perp}'}^2+\chi^2)(\mathbf{k}_{\perp}^2+\chi^2)}-
            	\frac{\mathbf{k}_{\perp}^2}{(\mathbf{k}_{\perp}^2+\chi^2)^2} \bigg)
            	\cos\bigg(\frac{\mathbf{k}_{\perp}^2+\chi^2}{2 z_+ p} \tau\bigg) \Bigg],
            \end{split}
            \label{eq.splitting_rate_cujet}
        \end{align}
        where $q$ stands for $q/\bar{q}$, $p$ is the momentum of the incoming fermion, $C^R_j$ is its appropriate Casimir operator and $\mathbf{k}_{\perp}$ is the transverse momentum of the photon with respect to the parent fermion. $N_f$ denotes the number of flavours and $\rho(T)$ is the density of the scattering centers in the QGP medium. The Debye mass, $m^2_D = 4\pi\alpha_s T^2 (1+N_f/6)$, is used in $\chi^2(T) = m_D^2(1-z_+)/2$ to regulate IR singularities and the LPM phase in the cosine term of Eq.~\ref{eq.splitting_rate_cujet}. The time dependence of the rate is explicit in $\tau$, interpreted as the time since the last splitting. Finally, the momentum fraction variables are given by $z = k/p,\;z_+ = z[1+\sqrt{1-(k_\perp/z p)^2}]/2$. In both gluon and photon radiation, the radiated particle is assumed to be collinear with the incoming parton.

     \vspace{2mm}
     \emph{Radiative energy loss in \martini} --- 
        Inelastic splittings in \martini\, are calculated within the \amy~\cite{amy} framework using techniques of Finite Temperature Field Theory. Similar to the LO-\dglv\, result above, \amy\, rates are also derived for asymptotically high temperatures and weak coupling limit but they differ on their view of the medium. \amy\, assumes that the hard parton is created in the infinite past and that it travels through an infinite QGP. Unlike LO-\dglv, the radiative rates of \amy\, do not have time dependence. 
        \begin{figure}
            \includegraphics[width=0.6\linewidth]{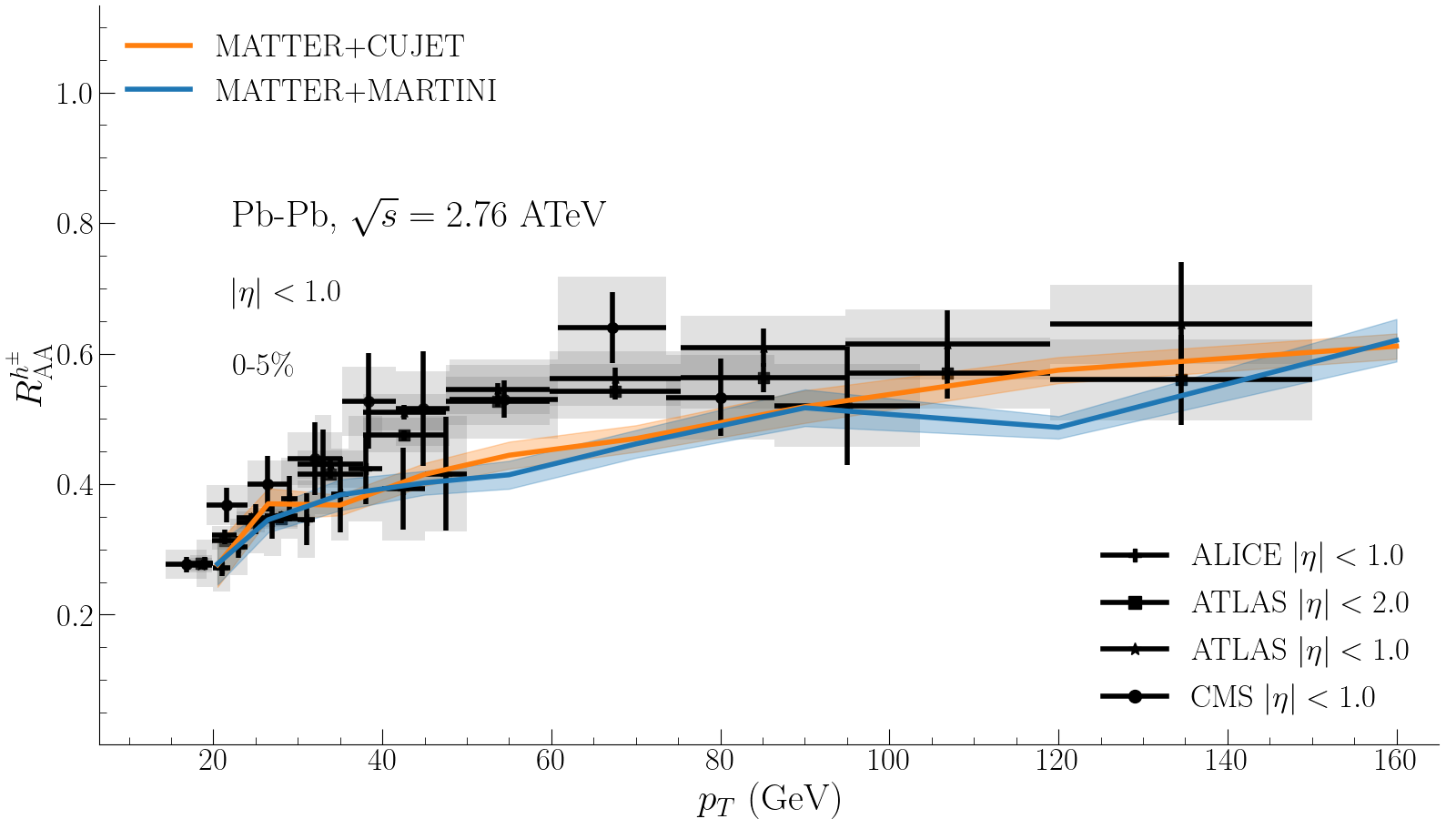}
            \begin{minipage}[b]{11pc}
            \begin{center} 
            \caption{
            Charged hadron nuclear modification factor, calculated using the composite models with \martini\, or \cujet\, for low virtuality energy loss. Data from ALICE, ATLAS and CMS Collaborations~\cite{experiment:chargedhadrons}.\label{fig:charged.hadron.raa}}
            \end{center}
            \end{minipage}
        \end{figure}
        The assumption of a weakly-coupled medium means the formation time of a radiation [$\tau_{\mathrm{form}}\propto (g^2_sT)^{-1}$] is much larger than the typical time of an elastic collision [$\tau_{\mathrm{elas}}\propto (g_sT)^{-1}$]. Thus during the radiation process, the incoming parton can receive an arbitrary number of kicks from the medium. Therefore to arrive at the bremsstrahlung rates, one has to re-sum an infinite set of ladder diagrams. Furthermore, the medium is assumed be at thermal equilibrium. Finally, \martini\, includes the gluon splitting to $q\bar{q}$ channel which \cujet\, does not.
        
     \vspace{2mm}
     \emph{Conversion Photons} --- 
        Conversion photons result from the passage of energetic fermions within a QGP medium. The channels through which the process occurs are QCD Compton scattering and $q\bar{q}$-annihilation. These are $t$-channel dominated processes and thus we make the \textit{collinear conversion} approximation~\cite{martini}, where no energy loss takes place during the process. \martini\, and \cujet\, implement this collinear approximation of jet-medium conversion photons. This results in a conversion photon spectrum that is directly proportional to the evolving hard fermion spectrum. To allow for a fair comparison between \cujet\, and \martini, we fix the strong coupling $\alpha_s=0.3$ for this photon channel.

\vspace{2mm}\emph{\textbf{Multi-probe study}} ---
    The models above are incorporated as models of the low-virtuality energy loss stage within a composite model. The hydrodynamic histories are generated using a \trento+Free-Streaming+\vishnu+\urqmd\, simulation with parameters fixed by Ref.~\cite{Bernhard:2019Nature}. For the jet simulations, we use the \jetscape\, workflow. Jets are produced using \pythia\, and evolve, on a parton-by-parton basis, using a composite model. At high-virtuality or in vacuum, \matter\, handles the splittings and energy loss and at low-virtuality, either \martini\, or \cujet\, govern the evolution. The partons freeze out of evolution when the local temperature falls below $T_c=160$ MeV or if the parton momentum is below $p_{\mathrm{cut}}=2$ GeV. Partons are hadronized at the end of the evolution, with jets clustered using the final state, stable hadrons. We use the parameter values fixed by the \jetscape\, Collaboration~\cite{JETSCAPE:pp,JETSCAPE:other} for all parameters of the model. 
    \begin{figure}
        \includegraphics[width=0.65\linewidth]{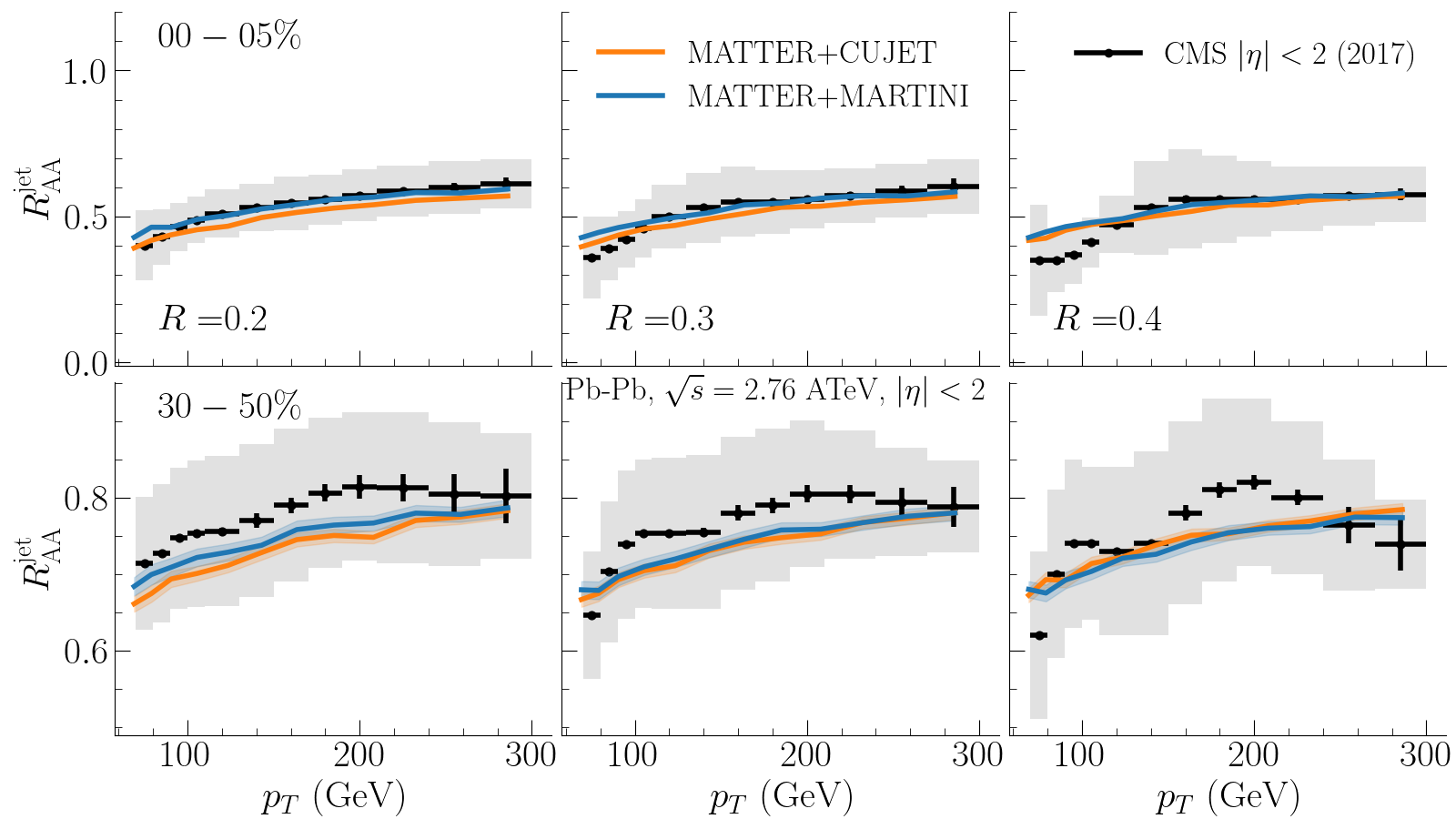}
        \begin{minipage}[b]{11pc}
        \begin{center} 
        \caption{
        Jet nuclear modification factor for inclusive jets at midrapidity. The calculations are shown for three values of cone size radius ($R$, columns) and two centrality classes (rows). Data from the CMS Collaboration~\cite{experiment:jets}. See text for details.\label{fig:jet.raa}}
        \end{center}
        \end{minipage}
    \end{figure}
    The exceptions are the \cujet\, and \martini-specific parameters, which we fit using the charged hadron \raa\, data at $0$-$5\%$ centrality class for \pbpb\, collisions at $2.76$ ATeV
    \begin{equation}
        R^{h^{\pm}}_{\mathrm{AA}}\left(p_T\right) = \frac{E\,d\sigma^{h^{\pm}}_{\nuclearcoll}/dp^3}{N_{\mathrm{bin.}}\,E\,d\sigma^{h^{\pm}}_{\pp}/dp^3}
        \label{eq:charged.hadron.raa}
    \end{equation}
    where $N_{\mathrm{bin.}}$ is the number of binary collisions. \autoref{fig:charged.hadron.raa} shows the fit result~\cite{Ourpaper}.
    
\begin{figure}
        \includegraphics[width=0.65\linewidth, height=0.25\linewidth]{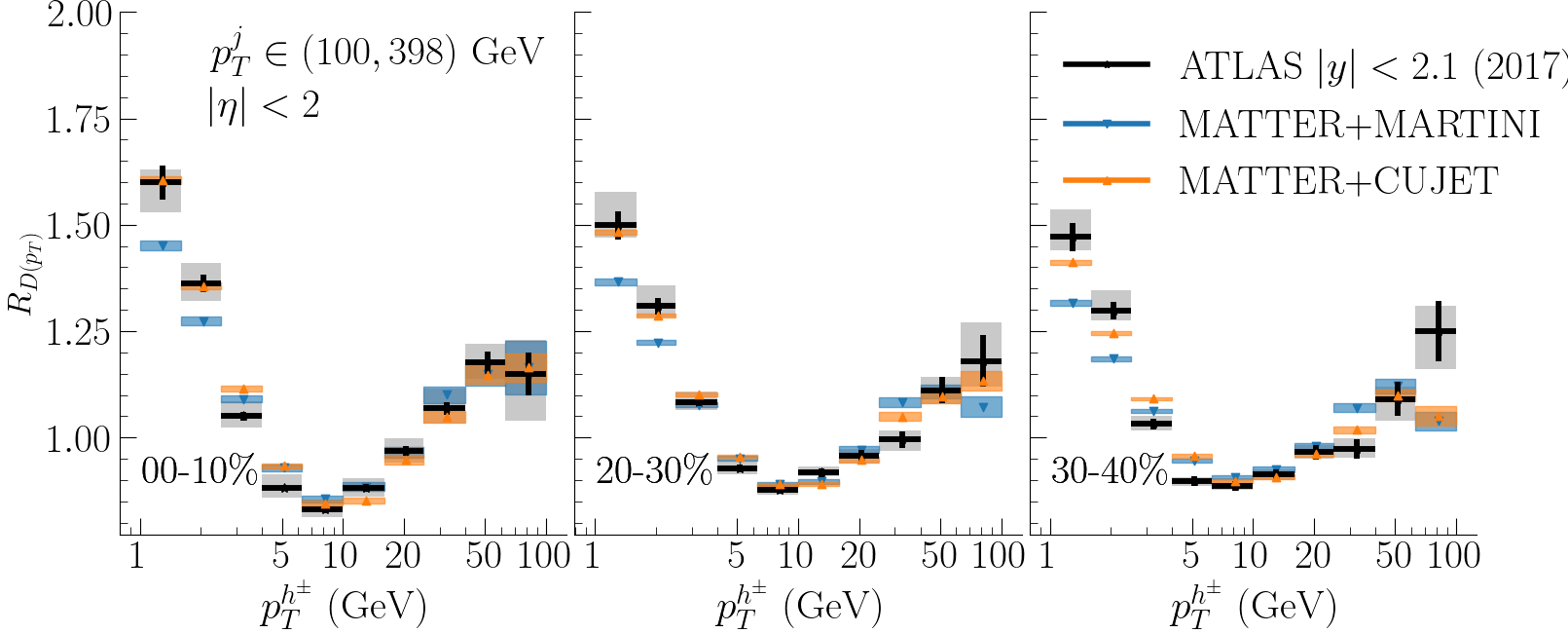}
        \begin{minipage}[b]{11pc}
            \begin{center}
                \caption{Jet fragmentation function ratio as a function of charged hadron $p_T$. Data from the ATLAS Collaboration~\cite{experiment:jetFF}. See text for details.\label{fig:jet.ff.ratios}}
            \end{center}
        \end{minipage}
    \end{figure}

    \autoref{fig:jet.raa} shows the results for inclusive jet \raa--defined in analogy to \autoref{eq:charged.hadron.raa}-- computed from the same simulations as \autoref{fig:charged.hadron.raa}. It is evident that the agreement with the data and among the models themselves is very good. However, we can observe a clear relative movement between the models in each centrality class, reflective of the difference between the models at rate level. \martini\, radiates more soft gluons relative to \cujet. These then get kicked away from the primary parton via elastic scatterings. Therefore studies of the jet cone dependence of the energy loss or jet substructure can be an illuminating probe of the models. The point is better illustrated by the jet fragmentation function (FF) ratio, measuring the medium modification to the $p_T$ distribution of charged hadrons within jets
    \begin{equation}
        R_{D\left(p_T\right)} = \frac{D_{\mathrm{AA}}\left(p_T\right)}{D_{\mathrm{pp}}\left(p_T\right)},\qquad
        D\left(p_T\right)= \frac{1}{N_{\mathrm{jet}}}\sum_{\mathrm{jets}}\sum_{p_{T,\mathrm{trk}}\in[p^{\mathrm{min}}_T,p^{\mathrm{max}}_T)}\frac{1}{p^{\mathrm{max}}_T-p^{\mathrm{min}}_T}.
    \label{eq:pp.FF.pT}
    \end{equation}
    \autoref{fig:jet.ff.ratios} shows the jet FF ratio calculated for the two composite models. As expected, there are systematic differences between the two models particularly for charged hadron $p_T< 5$ GeV. This is further evidence of the fact that \martini\, radiates more soft gluons than \cujet. These gluons then get pushed further and further out of the jet cone by elastic channels.
    As a complimentary probe, we study the contribution of jet-medium photons to the total direct photon spectrum by constructing the spectrum in three ways. The first spectrum is the sum of prompt, thermal and pre-equilibrium photons with no jet-medium input. The latter two spectra are taken from Ref.~\cite{Gale:multimessenger}. The prompt photon spectrum is calculated by a \pythia\, simulation.
    
\begin{figure}
        \includegraphics[width=0.68\linewidth, height=0.45\linewidth]{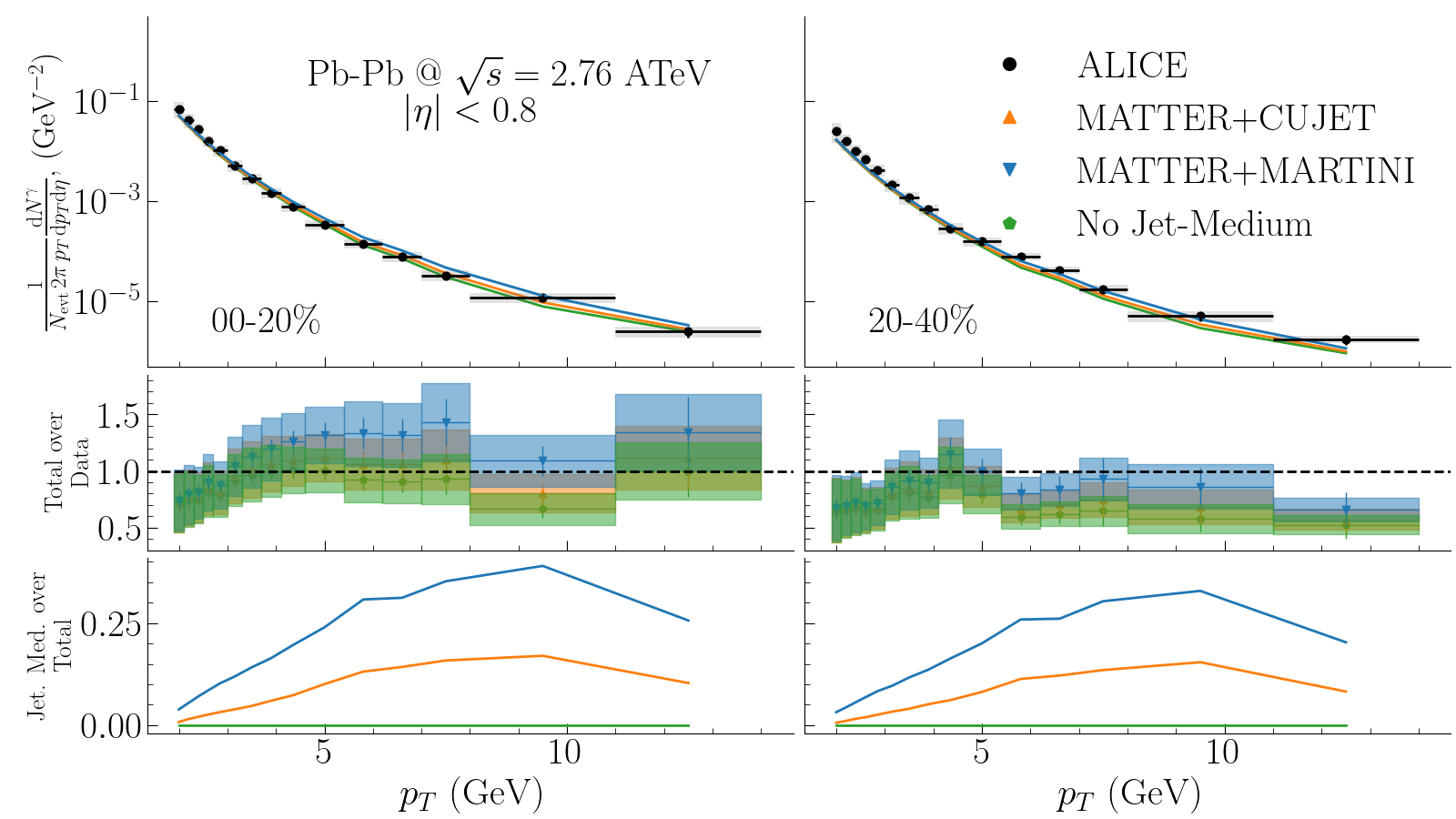}
        \begin{minipage}[b]{11pc}
        \begin{center} 
        \caption{
        Direct photon spectra for Pb-Pb collisions at $2.76$ ATeV. Top panel shows direct spectra against the data. Middle panel shows the ratio of the direct spectra to data. In the bottom panel, $r_{\mathrm{j.med.}}$ and $r_{\mathrm{other}}$ show the ratio of the jet-medium and all other channels to the total theory spectrum, respectively. Data from the ALICE Collaboration~\cite{experiment:photons}. See text for details.\label{fig:dir.gamma.spec}}
        \end{center}
        \end{minipage}
    \end{figure}

    The other two spectra are constructed by adding photons from either \matter+\martini\, or \matter+\cujet\, simulations to the first spectrum. \autoref{fig:dir.gamma.spec} shows the total direct photon spectrum, constructed for the three spectra specified above. 
    In the upper and middle panels of the figure, one can clearly see that the inclusion of jet-medium photons enhances the direct photon spectrum, particularly for the transverse momenta $p_T>4$ GeV. In this region, both models make a significant contribution to the total direct photon spectrum, though the contribution of \martini\, is much larger. The bottom panel of the figure shows the ratio of the calculated jet medium photon yield to the total theory yield. The jet-medium contribution is visibly significant and rises in importance between $4$-$12$ GeV, particularly for \martini. Contribution of jet-medium photons, particularly for \martini, shows strong $p_T$ dependence where they make up nearly half the photon yield at $p_T\approx 9$-$10$ GeV, double the jet-medium yield from \cujet. This is due \martini\, having a higher total rate of emission, for photons and gluons alike. Thus not only does \martini\, emit more photons, but also the spectrum of hard fermions evolved by \martini\, is softening faster than \cujet, resulting in more conversion photons for low and intermediate $p_T$. 

\vspace{2mm}\emph{\textbf{Conclusion}} ---
Energy loss of hard partons is an important signal of the creation of the quark-gluon plasma and a great probe of the medium. In this work we presented the results of the first multi-stage and multi-probe comparative study of \cujet\, and \martini\, using the \jetscape\, framework. The comparison includes charged hadron and jet nuclear modification factors as well as jet fragmentation function ratios. It was found~\cite{Ourpaper} that charged hadron \raa\, is not sensitive to the differences between the models while jet \raa\, shows a slight but systematic sensitivity. Jet fragmentation functions was found to be very sensitive, specially when considering the relatively soft hadrons within jets. As a part of this study, we presented the first calculation of photons from jet-medium interactions within a realistic plasma and a dynamic jet population for both \martini\, and \cujet. These photons are found to make a clear and significant contribution to the total direct photon yield in the intermediate $4\leq p_T \leq 12$ GeV. Specifically \martini\, sees their contribution as nearly half the total photon spectrum in this $p_T$ domain while \cujet\, places this contribution at approximately $25\%$. The conversion photon channel allows for the study of jet energy loss at a partonic level, without non-perturbative modifications from hadronization. This is further motivation for a comprehensive analysis of the jet-medium photons.

\vspace{2mm}\emph{\textbf{Acknowledgements}} ---
This work was funded in part by the Natural Sciences and Engineering Research Council of Canada, the U.S. Department of Energy, Office of Science, Office of Nuclear Physics, under Grant No.
DE-FG88ER41450, and Tsinghua University under Grant No. 53330500923. Computations were made on the B\'eluga, Graham and Narval computers managed by Calcul Qu\'ebec and by the Digital Research Alliance of Canada.


\begin{thebibliography}{99}
\bibitem{MATTER}
    A.~Majumder, Phys. Rev. C 88, 014909 (2013),
    A.~Majumder, J.~Putschke, Phys. Rev. C 93, 054909 (2016),
    A.~Majumder, S.~Cao, Phys. Rev. C 101, 024903 (2020).
\bibitem{martini} 
    B.~Schenke, C.~Gale, S.~Jeon, Phys. Rev. C 80 (2009) 054913 .
\bibitem{cujet}
    J.~Xu, A.~Buzzatti, M.~Gyulassy, JHEP 08 (2014) 063.
\bibitem{JETSCAPE:other}
    S.~Cao et al. (\jetscape\, Collaboration), Phys. Rev. C 104, 024905 (2021).
\bibitem{JETSCAPE:pp}
    A.~Kumar et al. (\jetscape\, Collaboration), Phys. Rev. C 102, 054906 (2020).
\bibitem{dglv}
    M.~Gyulassy, P.~Levai, I.~Vitev, Nucl.Phys.B 571 (2000) 197-233, Nucl.Phys.B 594 (2001) 371-419,
    M. Djordjevic, M. Gyulassy, Nucl.Phys.A 733 (2004) 265-298.
\bibitem{Zhang:2010hiv}
    H.~z.~Zhang, Z.~b.~Kang, B.~W.~Zhang and E.~Wang, Eur. Phys. J. C \textbf{67}, 445-454 (2010).
\bibitem{amy}
    P.~B.~Arnold, G.~D.~Moore, L.~G.~Yaffe, JHEP 11 (2001) 057,
    JHEP 12 (2001) 009, JHEP 06 (2002) 030.
\bibitem{Bernhard:2019Nature}
    J.~E.~Bernhard, J.~S.~Moreland, S.~Bass, Nat. Phys. 15, 1113–1117 (2019). 
\bibitem{Gale:multimessenger}
    C.~Gale, JF.~Paquet, B. Schenke, C.~Shen, Phys. Rev. C, 105 (2022) 1, 014909.
\bibitem{experiment:chargedhadrons}
    B.~Abelev et al. (ALICE Collaboration), Phys. Rev. Lett. 109, 252302 (2012),
    G.~Aad et al. (ATLAS Collaboration), JHEP 2015, 50 (2015),
    S.~Chatrchyan et al. (CMS Collaboration), Eur. Phys. J. C 72 (2012) 1945.
\bibitem{experiment:jets}
    V.~Khachatryan et al. (CMS Collaboration), Phys. Rev. C 96, 015202 (2017).
\bibitem{experiment:jetFF}
    M.~Aboud et al. (ATLAS Collaboration), Eur. Phys. J. C 77 (2017) 379.
\bibitem{experiment:photons}
    A.~Jaroslav et al. (ALICE Collaboration), Phys.Lett.B 754 (2016) 235.
\bibitem{Ourpaper}
    S.~Shi, R.~Modarresi Yazdi, C.~Gale, S.~Jeon, Phys. Rev. C 107, 034908 (2023)

\end{thebibliography}
\end{document}